# Data-driven rational function neural networks: a new method for generating analytical models of rock physics

Weitao Sun[1,2]

[1]School of Aerospace Engineering, Tsinghua University, Beijing, 100084, China
[2]Zhou Pei-Yuan Center for Applied Mathematics, Tsinghua University, Beijing, 100084, China
Corresponding author: Weitao Sun (sunwt@tsinghua.edu.cn)

**Key Points:**

- Rational function neural networks method is proposed to construct analytical velocity model directly from training data.

- The advantage is that it makes theoretical modeling very simple. Intensive physical analysis and mathematical derivation are avoided.

- Gassmann's equation is perfectly reproduced from a set of data with random noises by using this method.

**Abstract**

Seismic wave velocity of underground rock plays important role in detecting internal structure of the Earth. Rock physics models have long been the focus of predicting wave velocity. However, construction of a theoretical model requires careful physical considerations and mathematical derivations, which means a long research process. In addition, various complicated situations often occur in practice, which brings great difficulties to the application of theoretical models. On the other hand, there are many empirical formulas based on real data. These empirical models are often simple and easy to use, but may be not based on physical principles and lack a proper formulation of physics. This work proposed a rational function neural networks (RafNN) for data-driven rock physics modeling. Based on the observation data set, this method can deduce a velocity model which not only satisfies the actual data distribution, but also has a proper mathematical form reflecting the inherent rock physics. The Gassmann's equation, which is the most commonly used theoretical model relating bulk modulus of porous rock to mineral composition, porosity and fluid, is perfectly reconstructed by using data-driven RafNN. The advantage of this method is that only observational data sets are required to extract model equations, and no complex mathematical and physical processes are involved. This work opens up for the first time a new avenue on constructing analytical expression of velocity models using neural networks and field data, which is of great interest for exploring the heterogeneous structure of the Earth.

**Plain Language Summary**

Seismic wave travels through the Earth's interior, and its velocity depends on properties such as the mineral composition of the medium. It is an important task to establish a quantitative relationship (we call it a model) between the properties of the medium and the seismic wave velocity. Based on the model, the material composition and structure information of the earth interior can be deduced from the observed seismic wave velocity. Theoretical modeling has a good physical picture and mathematical form, but requires a high level of relevant expertise. In this paper, a data-driven rational function neural network method is proposed to obtain theoretical models directly from field data sets. The input of this method is a data set and the output is a speed model with analytic expressions. The main advantage of this method is that it can generate mathematical expressions directly from data. This process does not require traditional physical assumptions and mathematical derivation. Construction of analytical rock physics model becomes an easy work with the help of this method.

## 1 Introduction

Seismology has been a powerful tool in detecting the structure of the Earth's interior more than half a century. The principle of determining the location of discontinuities inside the earth is to identify the arrival time of the reflected and refracted signals of seismic waves. The depth of these discontinuities can be expressed as the product of the propagation time and velocity of seismic waves. The wave velocities depend on properties of underground material, such as mineral composition, compressibility, density, anisotropy, etc. Sharp jump in seismic velocities reflect a material discontinuity inside the Earth, providing valuable information of the structure of the planet.

Although the large scale structure (crust, mantel and core) have been known for a long time[*Adam M. Dziewonski and Anderson*, 1981; *A. M. Dziewonski and Gilbert*, 1971], the understanding of seismic heterogeneity in the Earth is still limited[*Ritsema and Lekić*, 2020]. In developing reference Earth model, anisotropic and anelastic velocity models are essential to satisfy the large amount of precise data, including normal mode, surface wave and body wave data. The anisotropy and regional inhomogeneity are significant in the first tens of kilometers and an average model cannot reflect the actual earth structure[*Adam M. Dziewonski and Anderson*, 1981]. The actual velocity gradients of underground formation are still unresolved.

Construction of velocity models has been a major research in geophysics and seismography. The goal of velocity modeling is to predicts bulk modulus changes in terms of mineral composition, porosity, solid and fluid properties. Iron content is the main factor affecting shear wave velocity of mantle material and may cause lateral variation in elastic properties of lower mantle[*Speziale et al.*, 2005]. First-principles modelling of $CaSiO_3$ perovskite has been used to reproduce the anti-correlation in bulk and shear-wave velocity[*S-i Karato and Karki*, 2001]. However, the theoretical model assumes that $CaSiO_3$ perovskite has a cubic structure and ignores the large 15% softening of shear wave velocity induced by temperature decrease[*Adams and Oganov*, 2006; *Stixrude and Lithgow-Bertelloni*, 2007]. A power law formula of seismic velocity changing with frequency and temperature has been proposed, and the power exponent parameter was roughly estimated for frequency between $10^{-8}$ Hz and 1Hz[*S Karato and Spetzler*, 1990]. A temperature derivative model was developed to explain temperature dependence of seismic-wave velocities[*S-i Karato*, 1993]. But the application of the model in the deep mantle is still uncertain. This is due to a limitation of the theoretical model, namely the assumption of spherically symmetric structure.

Empirical functions have also been proposed and are used to investigate the attenuation of seismic shear in upper mantle. The fitting parameter values were obtained based on the mechanical experiments data of olivine polycrystals [*Jackson et al.*, 2002]. Simple relationships were used to estimate shear velocity of silicate rocks from well logs, which are in agreement with Gassmann's equation for ideal model of sphere packing[*Castagna et al.*, 1985]. The shear and compressional velocities ratio have been established for different lithologies by using various empirical relationships[*Carroll*, 1969; *Greenberg and Castagna*, 1992; *Han et al.*, 1986; *Hossain et al.*, 2012; *Krief et al.*, 1990; *Krishna et al.*, 1989]. The importance of the empirical function is that it directly relates seismological velocity to experimental measurements. However, the formula is mostly used in limited zones and may produce inaccurate results for unconsidered dataset.

Recently, data-driven methods are developed to predict velocity in siliciclastic rocks[*Oloruntobi et al.*, 2019]. More interestingly, neural networks combined with data-driven method provides new possibilities in inverting S-wave velocity[*Fu et al.*, 2021; *Miah et al.*, 2021]. Neural networks has been used to estimate petrophysical models as inputs for full-waveform inversion[*Iturrarán-Viveros et al.*, 2021]. The main results obtained by training of neural networks is usually a profile of velocity vs. depth/time[*Fabien-Ouellet and Sarkar*, 2019; *Moya and Irikura*, 2010]. More recently, rational activation functions have been used as activation function of neural networks[*Boullé et al.*, 2020].

It is an exciting task to obtain analytical expressions of speed models directly from data sets. Extracting governing equations from noisy measurement data has been studied in nonlinear dynamical systems by using machine learning and sparse regression[*Brunton et al.*, 2016; *Champion et al.*, 2019]. The method provides interpretable models and enable deep discoveries in data-rich fields. However, the candidate functions learned from dataset are still limited to polynomial and trigonometric terms. The relationships between velocity and rock properties are often in form of rational functions. Examples of such model include Gassmann's equation[*Gassmann*, 1951], White model[*White*, 1975] and so on, which define velocity or bulk modulus of underground rocks as a rational function of mineral modulus, porosity and pore fluid stiffness. The rational function neural networks (RafNN) in this work is inspired by learning rational function of velocity models directly from large amount of data sets.

Rational function, which is defined as the ratio of two polynomial functions, represents a general mathematical representation and can achieve a wide range of linear and nonlinear system simulation with arbitrary precision. Moreover, the rational function is easy to operate mathematically and provides a convenient model description method[*Chen and Billings*, 1989]. Rational function has difficulties in model identification and parameter estimation[*Zhu and Billings*, 1994]. Least squares regression has been used to determine nonlinear rational models[*Billings and Zhu*, 1991; *Zhu*, 2005]. An error back propagation algorithm was proposed to improve the model parameter estimation[*Zhu*, 2003]. Neural computing algorithms is also applied in determining the parameters of rational function model[*Liu et al.*, 2020].

This work aims to develop rational function neural networks that extracts analytical expression of rock physics model from field datasets, rather than providing a 1-D or 2-D velocity profile. The main advantage of the data-driven RafNN modeling is that it can generate mathematical expressions directly from data. This process does not require traditional physical assumptions and mathematical derivation. Such a data-driven rock physics model not only satisfied the observed training data, but also is efficient in computing velocities with mineral/rock properties.

There are important differences between data-driven velocity model base on RafNN and traditional data regression/fitting methods. First of all, data fitting methods generally use relatively simple traditional inherent models (such as linear, quadratic, Gaussian, power-law models, etc.), while RafNN allows to deal with more common types of mathematical models, or even a mixture of existing models. Therefore, this method is increasingly suitable for the application of science and engineering data intensive systems. Secondly, data-driven modeling methods are used to discover physical systems that describe and predict the behavior of models. Traditional data fitting methods pay more attention to the morphology and classification of data, while data-driven modeling methods are based on mathematical forms with physical meaning, which can explain the existing data, provide effective predictions, and be verified by future experimental data.

## 2 Rational functional neural networks

A complex nonlinear function can be expressed in the following rational form

$$y(\phi) = \frac{\sum_{j_1=0}^{p}\sum_{j_2=0}^{p}\cdots\sum_{j_n=0}^{p} c_{j_1 j_2 \cdots j_n} x_1^{j_1} x_2^{j_2} \cdots x_n^{j_n}}{\sum_{k_1=0}^{q}\sum_{k_2=0}^{q}\cdots\sum_{k_l=0}^{q} d_{k_1 k_2 \cdots k_n} x_1^{k_1} x_2^{k_2} \cdots x_l^{k_l}} + o(\phi), \quad (1)$$

where $\phi$ is the independent variable and $x_i(\phi)$ ($i=1,...,n$) is a function of $\phi$. $c$ and $d$ are coefficients of the polynomial $x_1^{j_1} x_2^{j_2} \cdots x_n^{j_n}$ and $x_1^{k_1} x_2^{k_2} \cdots x_n^{k_n}$, $o(\phi)$ is an error function. $p$ and

$q$ indicate the order of the numerator and denominator polynomials.

Taking Gassmann's equation as an example, the volumetric modulus $K_{sat}(K_d,\phi;K_m,K_f)$ of fluid-containing porous media is a function of the bulk modulus of dry skeleton ($K_d$), porosity $\phi$, modulus of solid material ($K_m$) and bulk modulus of pore fluid $K_f$ [Gassmann, 1951].

$$K_{sat} = K_d + \frac{\left(1-\dfrac{K_d}{K_m}\right)^2 K_f}{\phi + \left(1 - \dfrac{K_d}{K_m} - \phi\right)\dfrac{K_f}{K_m}} \qquad (2)$$

$$= \frac{-K_d K_f K_m + K_m^2 K_f + K_d \phi K_m^2 - K_d K_f K_m \phi}{-K_d K_f + K_m K_f + \phi K_m^2 - \phi K_m K_f}$$

By defining an independent variable vector as $\mathbf{x} = (K_d, \phi, K_m, K_f)$, one can find that the highest order of the numerator polynomial is $p = 4$, and the highest order of the denominator polynomial is $q = 3$. The Gassmann's equation can be rewritten in a more general form as

$$K_{sat} = \frac{\sum\limits_{\substack{j_1=0 \\ (j_1+\ldots+j_n=4)}}^{4}\sum\limits_{j_2=0}^{4}\sum\limits_{j_3=0}^{4}\sum\limits_{j_4=0}^{4} c_{j_1 j_2 j_3 j_4} K_d^{j_1} \phi^{j_2} K_m^{j_3} K_f^{j_4}}{\sum\limits_{\substack{k_1=0 \\ (k_1+\ldots+k_n=3)}}^{3}\sum\limits_{k_2=0}^{3}\sum\limits_{k_3=0}^{3}\sum\limits_{k_4=0}^{3} d_{k_1 k_2 k_3 k_4} K_d^{k_1} \phi^{k_2} K_m^{k_3} K_f^{k_4}} + o(\phi), \qquad (3)$$

where, the coefficients are $c_{1011} = -1$, $c_{0021} = 1$, $c_{1120} = 1$, $c_{1111} = -1$, $d_{1001} = -1$, $d_{0011} = 1$, $d_{0120} = 1$, $d_{0111} = -1$. Other coefficients are all zeros.

Write the rational approximation equation of any function in vector form, and get

$$y(\phi) = \frac{\Xi_N \mathbf{c}}{\Xi_D \mathbf{d}} + o(\phi), \qquad (4)$$

where $\Xi_N$ is a matrix composed of polynomials of functions $x(\phi)$

$$\Xi_N = \begin{bmatrix} 1 & x_1(\phi_1) & \ldots & x_n(\phi_1) & x_1^2(\phi_1) & x_1(\phi_1)x_2(\phi_1) & \ldots & x_n^2(\phi_1) & \ldots & x_1^p(\phi_1) & \ldots & x_n^p(\phi_1) \\ 1 & x_1(\phi_2) & \ldots & x_n(\phi_2) & x_1^2(\phi_2) & x_1(\phi_2)x_2(\phi_2) & & x_n^2(\phi_2) & & x_1^p(\phi_2) & \ldots & x_n^p(\phi_2) \\ \ldots & & & & \ldots & & & & & & & \\ 1 & x_1(\phi_m) & \ldots & x_n(\phi_m) & x_1^2(\phi_m) & x_1(\phi_m)x_2(\phi_m) & & x_n^2(\phi_m) & & x_1^p(\phi_m) & \ldots & x_n^p(\phi_m) \end{bmatrix}. \qquad (5)$$

The numerator coefficient matrix is

$$\mathbf{c} = \begin{bmatrix} c_{0000} \\ c_{1000} \\ \dots \\ c_{000p} \end{bmatrix}. \tag{6}$$

The denominator coefficient matrix is

$$\mathbf{d} = \begin{bmatrix} d_{0000} \\ d_{1000} \\ \dots \\ d_{000p} \end{bmatrix}. \tag{7}$$

The above rational function can be described by the rational function neural network (RafNN). RafNN has three parts: input layer, hidden layer and output layer. The input layer consists of polynomial functions of numerator and denominator. The hidden layer is composed of the linear combination of the numerator input layer and the denominator input layer respectively. The output layer consists of the ratio of the numerator and denominator terms. The weighted network connecting the input layer and the hidden layer is polynomial coefficients $\mathbf{c}$ and $\mathbf{d}$ (Figure 1). After training and learning process, the coefficients of the weighted network make the output layer function value consistent with the expected value.

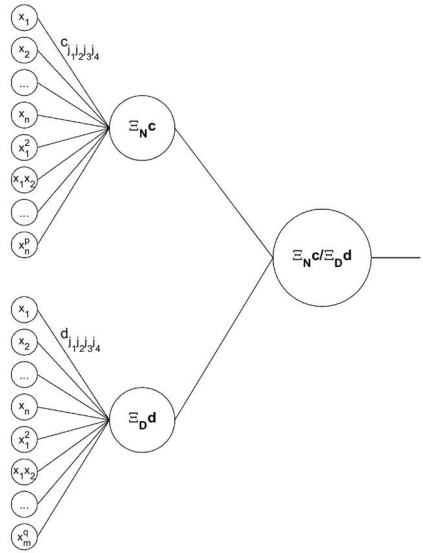

Figure 1 Schematic diagram of rational function neural network (RafNN) unit

Since the input layer of RafNN contains many polynomials, the weight coefficients after training should try to meet the requirement of sparsity, i.e., to approximate the real function value with as few terms as possible.

The neural network cost function is defined as follows

$$E(\phi) = \frac{1}{2}\left\| y(\phi) - \frac{\Xi_N \mathbf{c}}{\Xi_D \mathbf{d}} \right\|_2^2 + \lambda_1 \|\mathbf{c}\|_1 + \lambda_2 \|\mathbf{d}\|_1. \tag{8}$$

The first term represents the approximation error of the neural network to a nonlinear function

$y(\phi)$. The second and third terms are norms of the weight coefficients of RafNN, which aim to make the network as sparse as possible. $\lambda_1$ and $\lambda_2$ are the coefficients controlling sparse constraints.

The training process of RafNN includes forward propagation and error feedback process. The forward propagation of the network is calculated by the following formula

$$\Xi_N \mathbf{c}(i+1) = \sum_{j_1=0}^{p}\sum_{j_2=0}^{p}\cdots\sum_{j_n=0}^{p} c_{\mathbf{j}}(i)\mathbf{x}^{\mathbf{j}}(i+1), \quad (j_1+\ldots+j_n=p) \tag{9}$$

$$\Xi_D \mathbf{d}(i+1) = \sum_{k_1=0}^{q}\sum_{k_2=0}^{q}\cdots\sum_{k_l=0}^{q} d_{\mathbf{k}}(i)\mathbf{x}^{\mathbf{k}}(i+1), \quad (k_1+\ldots+k_n=q) \tag{10}$$

$$\hat{y}(i+1) = \frac{\Xi_N \mathbf{c}(i+1)}{\Xi_D \mathbf{d}(i+1)}, \tag{11}$$

$$o(i+1) = y(i+1) - \hat{y}(i+1). \tag{12}$$

Here $i, i+1$ represents the subscript corresponding to each sample data in the training data set.

Here an error back propagation (BP) rational functional neural network algorithm is proposed. The error feedback of RafNN is achieved by the following formula

$$\frac{\partial E(i+1)}{\partial c_{\mathbf{j}}(i+1)} = -[y(i+1)-\hat{y}(i+1)]\frac{\partial(y(i+1)-\hat{y}(i+1))}{\partial(\Xi_N \mathbf{c})}\frac{\partial(\Xi_N \mathbf{c})}{\partial c_{\mathbf{j}}} + \lambda_1 \frac{\partial \|\mathbf{c}\|_1}{\partial c_{\mathbf{j}}},$$
$$= -o(i+1)\frac{\mathbf{x}^{\mathbf{j}}}{\Xi_D \mathbf{d}(i+1)} + \lambda_1 \frac{c_{\mathbf{j}}(i+1)}{\sqrt{c_{\mathbf{j}}^2(i+1)}} \tag{13}$$

$$\frac{\partial E(i+1)}{\partial d_{\mathbf{j}}(i+1)} = -[y(i+1)-\hat{y}(i+1)]\frac{\partial(y(i+1)-\hat{y}(i+1))}{\partial(\Xi_D \mathbf{d})}\frac{\partial(\Xi_D \mathbf{d})}{\partial d_{\mathbf{k}}} + \lambda_2 \frac{\partial \|\mathbf{d}\|_1}{\partial d_{\mathbf{k}}},$$
$$= o(i+1)\frac{\Xi_N \mathbf{c}}{(\Xi_D \mathbf{d})^2}\mathbf{x}(i+1)^{\mathbf{k}} + \lambda_2 \frac{d_{\mathbf{k}}(i+1)}{\sqrt{d_{\mathbf{k}}^2(i+1)}} \tag{14}$$

$$c_{\mathbf{j}}(i+1) = c_{\mathbf{j}}(i) - \eta \frac{\partial E(i+1)}{\partial c_{\mathbf{j}}(i+1)}, \tag{15}$$

$$d_{\mathbf{k}}(i+1) = d_{\mathbf{k}}(i) - \eta \frac{\partial E(i+1)}{\partial d_{\mathbf{k}}(i+1)}. \tag{16}$$

Here $i = 1,\ldots L$, L is the length of training data. $c_{\mathbf{j}}$ is $c_{j_1 j_2 \ldots j_n}$, $j = 1,2,\ldots,p$ and $\sum_{i=1}^{n} j_i = p$.

$d_{\mathbf{k}}$ is $d_{k_1 k_2 \ldots k_m}$, $k = 1,2,\ldots,q$ and $\sum_{i=1}^{m} k_i = q$. Here $p$ and $q$ are the order of the rational function.

$\eta$ is learning rate.

The operations of RafNN are given below.

Step 1: Generate RafNN input layer data.

According to the training data variable $\mathbf{x} = \{x_1, x_2, ..., x_n\}$ and the order of the rational formula, the input layer data $\Xi_N = [1, \Xi_1, ..., \Xi_p]$ and $\Xi_D = [1, \Xi_1, ..., \Xi_q]$ of the neural network is generated, where $\Xi_1 = [x_1, x_2, ..., x_n]$, $\Xi_2 = [x_1^2, x_1x_2, ..., x_1x_n, x_2^2, x_2x_3, ..., x_n^2]$ and $\Xi_p = [x_1^p, x_1^{p-1}x_2, x_1^{p-1}x_3, ..., x_1^{p-1}x_n, x_1^{p-2}x_2^2, x_1^{p-2}x_3^2, ..., x_n^p]$. Here $x_i(\phi)$ is a function of our basic variable $\phi$.

Step 2: Initialize the RafNN weight coefficients.

The network weight coefficients $c_\mathbf{j}$ and $d_\mathbf{k}$ are uniformly distributed random numbers. The random number has a 0 mean and a small variance.

Step 3: Training RafNN parameters.

for m=0 to M (Training iterative cycle)

for i=1 to L (Training sample length cycle)

(1) Forward propagation calculation

$$R_N(i+1) = \sum_{j_1=0}^{p}\sum_{j_2=0}^{p}...\sum_{j_n=0}^{p} \sigma(|c_\mathbf{j}(i)| - \varepsilon)\mathbf{x}^\mathbf{j}(i+1), \quad (17)$$
$$(j_1+...+j_n=p)$$

$$R_D(i+1) = \sum_{k_1=0}^{q}\sum_{k_2=0}^{q}...\sum_{k_l=0}^{q} \sigma(|d_\mathbf{k}(i)| - \varepsilon)\mathbf{x}^\mathbf{k}(i+1), \quad (18)$$
$$(k_1+...+k_n=q)$$

$$\hat{y}(i+1) = \frac{R_N(i+1)}{R_D(i+1)}, \quad (19)$$

$$o(i+1) = y(i+1) - \hat{y}(i+1), \quad (20)$$

Here $\sigma(.)$ is Heaviside function, which means that when the absolute value of the coefficient $c_\mathbf{j}$ is less than the value $\varepsilon$, the coefficient is given a value of 0.

(2) Error back propagation.

$$\delta c_\mathbf{j} = -\frac{o(i+1)}{R_D(i+1)}\mathbf{x}^\mathbf{j} + \lambda_1 \frac{c_\mathbf{j}(i+1)}{\sqrt{c_\mathbf{j}^2(i+1)}}, \quad (21)$$

$$\delta d_\mathbf{k} = o(i+1)\frac{R_N}{R_D^2}\mathbf{x}(i+1)^\mathbf{k} + \lambda_2 \frac{d_\mathbf{k}(i+1)}{\sqrt{d_\mathbf{k}^2(i+1)}}, \quad (22)$$

$$c_\mathbf{j}(i+1) = c_\mathbf{j}(i) - \eta\,\delta c_\mathbf{j}, \quad (23)$$

$$d_k(i+1) = d_k(i) - \eta \, \delta d_k. \tag{24}$$

End (end of training sample length cycle)
End (end of training iteration loop)

## 3 Model structure and parameter determination for data-driven RafNN

The mathematical expression of neural network for data-driven velocity modeling contains a large number of functions to be determined. Parameter estimation is simplified when one knows which function terms to fit. For example, in the process of deriving Gassmann's equation from data-driven neural network, the rational function contains 69 molecular function terms and 34 denominator function terms. In the process of neural network initialization, random initialization values are assigned to all function coefficients. Despite the large number of coefficients, prior knowledge of the mathematical expression of the rock physical model can be used because the data training set is known to satisfy the Gassmann's equation. In particular, we know that the four numerator terms and the four denominator terms should have the form $K_d K_f K_m, K_m^2 K_f, K_d \phi K_m^2, K_d K_f K_m \phi$ and $K_d K_f$, $K_m K_f$, $\phi K_m^2$, $\phi K_m K_f$. The true values of these coefficients are known in the approximate range of (-1, 1, 1, 1). Therefore, prior knowledge is introduced into the model coefficient initialization. The coefficients of these corresponding function terms can be changed near the true value to speed up the convergence process of the model.

However, in many practical applications, the prior knowledge of model structure cannot be given. As a result, the number of functional terms in nonlinear models will increase rapidly, and model users will have to search a wide range of model spaces to match the characteristics of the data. And the final model expression may take many forms. Therefore, it is very important to find a concise and accurate rational function expression of rock physics model.

## 4 Reconstruction of the Gassmann's equation from logging data

Different rocks in nature have specific properties and may follow different relations between velocity and rock physics parameters (bulk modulus, density and porosity, etc.). Velocity models satisfying all kinds of relations form a model space. Theoretical models (such as Gassmann's equation[*Gassmann*, 1951], Biot equation[*Biot*, 1956], White model[*White*, 1975], Johnson model[*Johnson*, 2001], BISQ model[*Dvorkin and Nur*, 1993], BIPS model[*Sun*, 2021], etc.) are subset in the model space and have the ability to describe specific types of behaviors if corresponding rocks. However, none of these models can completely cover the whole model space. In practice, there are always some data that cannot be explained by existing models.

The rock physics model based on RafNN is a data-based modeling method. As long as the relevant data set is given, the corresponding rock physics model can be obtained through neural network. The model expressions obtained in this way can well describe the rock physical laws followed by these data. At the same time, the prediction results of the neural network are interpretable. It can be thought that RafNN is a method of automatic modeling of rock physical. This method may have the potential to replace the "theoretical modeling" which requires intensive mathematics, physics and rock physics background knowledge. In addition, RafNN achieves fast and adaptive "data-driven modeling" for any target region with massive real data.

Gassmann's equation has been derived based on sphere packing model of a single type of mineral particle[*Gassmann*, 1951]. The bulk modulus is a function of mineral and fluid parameters (Equation 2). Driven by the data obeying Gassmann's equation, a rock physics model with the expression of same mathematical form is obtained by training RafNN. The training data set and test data set were generated by Gassmann's equation. Random noise perturbation of 1% of the amplitude was added to all data sets. The porosity ranges from 0.01 to 0.4. The skeleton is made of sandstone with bulk and shear moduli as $K_m = 37$ GPa, $G_m$ =44GPas. The pore fluid is water ( $K_f$ = 2.15GPa). The bulk modulus of dry skeleton is calculated by the formula proposed by Toksoz[*Toksoz et al.*, 1976]

$$K_d = \frac{K_m(1-\phi)}{1+3\phi K_m / G_m}. \tag{25}$$

Let the parameter vector be $\mathbf{x} = \{K_d, \phi, K_m, K_f\}$, the order of numerator rational functions is $p = 4$, and the highest order of the denominator function is $q = 3$. Then the input layer data of numerator terms consists 69 terms { $K_d, \phi, K_m, K_f$, $K_d^2, K_d\phi, K_dK_m, K_dK_f, \phi^2, \phi K_m, \phi K_f$, $K_m^2, K_mK_f, K_f^2, \ldots, K_f^4$ }. The denominator input layer contains 34 terms.

The threshold value of the network weight coefficient is set as =0.1. The number of network training cycles is 1000. The network has been independently trained for 35 times. Among the 35 rational neural networks whose mean square error of cost function is less than 1, the number of numerator and denominator of rational functions converges to 4 respectively as the error decreases (Figure 2). This is consistent with the expression of Gassmann's equation.

A RafNN with the smallest mean square error is used to obtain the rock physics model. The mathematical expression is:

$$K_{sat} = \frac{-0.80003 K_d K_f K_m + 0.79999 K_m^2 K_f + 0.79999 K_d \phi K_m^2 - 0.79999 K_d K_f K_m \phi}{-0.79997 K_d K_f + 0.79993 K_m K_f + 0.79999 \phi K_m^2 - 0.79998 \phi K_m K_f} \tag{26}$$

Although the coefficients obtained are different from those in Equation (2), the relationship between coefficients is still consistent with Gassmann's equation. By normalizing the coefficients, one gets the following expression

$$K_{sat} = \frac{-1.0001 K_d K_f K_m + K_m^2 K_f + K_d \phi K_m^2 - K_d K_f K_m \phi}{-K_d K_f + 0.99996 K_m K_f + \phi K_m^2 - \phi K_m K_f} \tag{27}$$

The predicted coefficients are highly consistent with the results of Gassmann's equation (Equation (2)). Six of the eight rational function coefficients are exactly the same. The errors of the other two coefficients from the Gassmann's equation coefficients are 0.01% and 0.004%, respectively. Since the above equation is directly extracted from data set by using RafNN, it can be seen that the traditional theoretical rock physics model can be perfectly reconstructed from the data-driven RafNN.

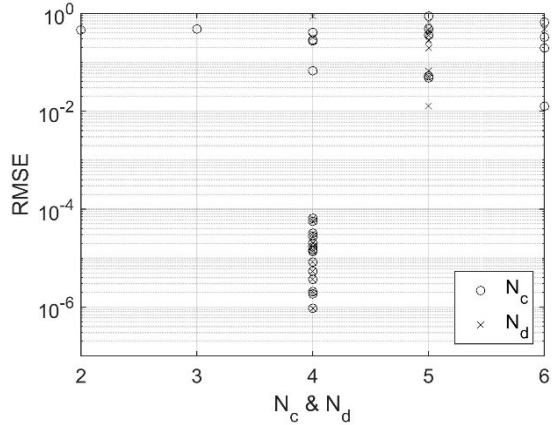

Figure 2 The number of numerator terms (Nc) and the number of denominator terms (Nd) both converge to 4 as the training error of RafNN decreases.

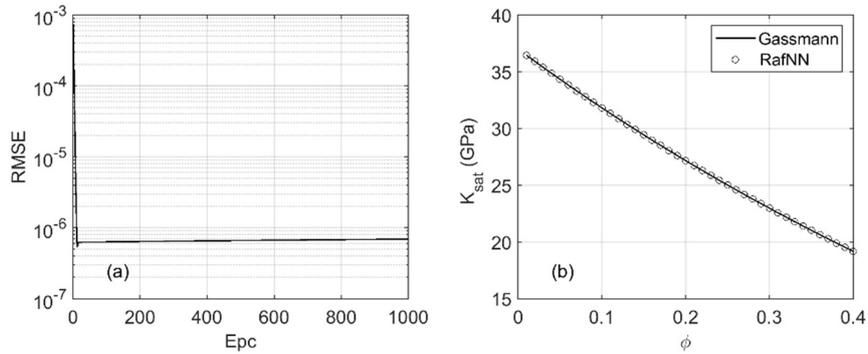

Figure 3 Training results of RafNN on data set of Gassmann's equation. (a) The root mean square error of the RafNN cost function vs. training epoch. (b) The predicted bulk moduli vs. porosity of water saturated rock are perfectly consistent with the theoretical values of Gassmann's equation.

In the 35 independent trainings, different modeling expressions other than Gassmann's equation are also obtained. The Table 1 and Table 2 in Appendix show coefficients of the rational functions in the numerator and denominator. The coefficients of final rational function model are different. However, these coefficients are still highly consistent with the coefficients of Gassmann's equation, with an error of less than 0.03% (Figure 4).

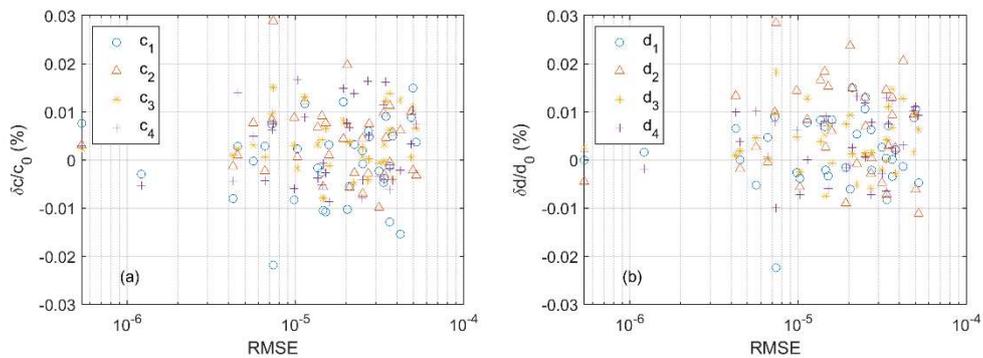

Figure 4  Coefficient error distribution of rock physics models obtained by RafNN. (a) The coefficient errors of the numerator terms are all within 0.03%. (b) The coefficient errors of denominator terms are all within 0.03%

In addition, for the data sets that cannot be explained by the theoretical model, this method can still obtain the corresponding mathematical expressions. The expression interprets and predict the corresponding observed data. Such unexpected analytical functions may have not been studied by the literature. However, these model expressions, which may be overlooked, may yield much more information about the structure and properties of the target rock.

In other words, the velocity model obtained by theoretical modeling is a special case of the data-driven RafNN modeling method. The different mathematical expressions obtained by RafNN constitute a more general set of models in the velocity model space, which may not be found by ordinary theoretical modeling methods. Therefore, the RafNN are helpful and enlightening to the development of theoretical modeling work.

## 5 Conclusion

A data-driven rational function neural networks (RafNN) method is established. Compared with traditional theoretical model construction process, data-driven RafNN modeling does not require theoretical assumptions and mathematical derivation. Building a model with proper mathematical form becomes fast and has high adaptability. Based on the RafNN and the field data set, one can build analytical equations of rock physics model for different kinds of rocks. Starting from the data set satisfying Gassmann's equation, the mathematical form of Gassmann's equation is perfectly reproduced. In addition, more models of undiscovered mathematical forms were obtained that are different from the standard model (such as Gassmann's equation). These models are derived entirely from data, and the mathematical expressions may have rich physical implications that have never been discovered before. The RafNN is a powerful way to construct analytical models from real data set, which cannot be fully expressed by the traditional models.

## Acknowledgments

This study was supported by the National Natural Science Foundation of China (Grant no. 41874137, Grant no. 42074144). The pore structure data associated with Figure 2, 3 and 4 in this article can be accessed at the website (https://www.researchgate.net/profile/Weitao-Sun-4/research).## References

Adams, D. J., and A. R. Oganov (2006), Ab initio molecular dynamics study of CaSiO3 perovskite at PT conditions of Earth's lower mantle, *Physical Review B*, *73*(18), 184106, doi:10.1103/PhysRevB.73.184106.
Billings, S. A., and Q. M. Zhu (1991), Rational model identification using an extended least-squares algorithm, *International Journal of Control*, *54*(3), 529-546, doi:10.1080/00207179108934174.
Biot, M. A. (1956), Theory of Propagation of Elastic Waves in a Fluid-Saturated Porous Solid .1. Low-Frequency Range, *J Acoust Soc Am*, *28*(2), 168-178, doi:Doi 10.1121/1.1908239.
Boullé, N., Y. Nakatsukasa, and A. Townsend (2020), Rational neural networks, in *NeurIPS 33*, edited.

# Appendix

Figure 1 The root mean square error of RafNN modeling and the coefficients of numerator rational function terms

| RMSE | Kd*Ks*Kf | Ks*Ks*Kf | Kd*phi*Ks*Ks | Kd*phi*Ks*Kf |
| --- | --- | --- | --- | --- |
| 5.358292e-07 | 1.0001 | -1 | -1 | 1 |
| 1.214217e-06 | 0.99995 | -1.0001 | -1 | 0.99993 |
| 4.261858e-06 | 0.99985 | -0.99992 | -0.99994 | 0.99989 |
| 4.518667e-06 | 1 | -1 | -1 | 1.0001 |
| 5.624624e-06 | 1 | -1.0001 | -1.0001 | 1.0001 |
| 6.584518e-06 | 0.99998 | -0.99993 | -0.99996 | 0.99991 |
| 7.260416e-06 | 0.99998 | -0.99999 | -1 | 0.99997 |
| 7.385096e-06 | 1 | -1.0005 | -1.0004 | 1.0003 |
| 9.818503e-06 | 0.99994 | -1.0001 | -1.0001 | 0.99997 |
| 1.024647e-05 | 1.0001 | -1 | -1 | 1.0002 |
| 1.133560e-05 | 1 | -1 | -1.0001 | 1 |
| 1.355369e-05 | 0.9999 | -0.99999 | -0.99995 | 0.99988 |
| 1.436550e-05 | 0.99991 | -1 | -0.99997 | 0.99988 |
| 1.457189e-05 | 0.99992 | -0.99996 | -0.99994 | 1 |
| 1.510996e-05 | 0.99993 | -1.0001 | -1.0001 | 1 |
| 1.581476e-05 | 0.99995 | -0.99993 | -0.9999 | 0.99983 |
| 1.926383e-05 | 1.0001 | -1.0001 | -1.0001 | 1.0002 |
| 2.031645e-05 | 0.99996 | -1.0003 | -1.0001 | 1.0001 |
| 2.097353e-05 | 0.9998 | -0.99979 | -0.99989 | 0.99992 |
| 2.233195e-05 | 0.99998 | -0.99992 | -0.9999 | 1.0001 |
| 2.493212e-05 | 0.99991 | -0.99994 | -0.9999 | 0.99981 |
| 2.511214e-05 | 0.99986 | -0.9998 | -0.9999 | 0.99983 |
| 2.708247e-05 | 1 | -0.99991 | -0.9999 | 1.0001 |
| 2.737077e-05 | 1.0001 | -1.0001 | -1 | 1.0001 |
| 3.145850e-05 | 0.99995 | -0.99988 | -0.99997 | 0.99994 |
| 3.335238e-05 | 0.99995 | -1.0001 | -0.99999 | 1.0001 |
| 3.367505e-05 | 1 | -1.0001 | -1.0001 | 1 |
| 3.443855e-05 | 1 | -0.99998 | -1.0001 | 1.0001 |
| 3.625296e-05 | 0.99991 | -1 | -1.0001 | 1 |
| 3.628892e-05 | 0.99999 | -1.0001 | -1.0001 | 0.99994 |
| 3.784736e-05 | 1 | -1 | -0.99994 | 0.99994 |
| 4.201730e-05 | 0.99986 | -1.0001 | -1.0001 | 0.99999 |
| 4.874307e-05 | 1 | -1 | -0.99992 | 0.99995 |
| 4.996208e-05 | 1 | -0.99987 | -1 | 1 |
| 5.204527e-05 | 1.0001 | -1 | -1.0001 | 1.0001 |

Table 2 The root mean square error of RafNN modeling and the coefficient of denominator rational function terms

| RMSE | Kd*Ks*Kf | Ks*Ks*Kf | Kd*phi*Ks*Ks | Kd*phi*Ks*Kf |
|---|---|---|---|---|
| 5.358292e-07 | 1 | -0.99996 | -1 | 1 |
| 1.214217e-06 | 1 | -1.0001 | -1.0001 | 0.99996 |
| 4.261858e-06 | 1 | -1.0001 | -0.99994 | 1 |
| 4.518667e-06 | 1 | -0.99998 | -1 | 1 |
| 5.624624e-06 | 1 | -1.0001 | -1.0001 | 1.0002 |
| 6.584518e-06 | 1 | -0.99995 | -0.99996 | 0.99996 |
| 7.260416e-06 | 1 | -1 | -1 | 0.99999 |
| 7.385096e-06 | 1 | -1.0005 | -1.0004 | 1.0001 |
| 9.818503e-06 | 1 | -1.0002 | -1.0001 | 1.0001 |
| 1.024647e-05 | 1 | -0.99998 | -1 | 0.99997 |
| 1.133560e-05 | 1 | -1 | -1.0001 | 0.99992 |
| 1.355369e-05 | 1 | -1.0001 | -0.99996 | 0.99999 |
| 1.436550e-05 | 1 | -1.0001 | -0.99998 | 1 |
| 1.457189e-05 | 1 | -1 | -0.99995 | 1.0001 |
| 1.510996e-05 | 1 | -1.0002 | -1.0001 | 0.99997 |
| 1.581476e-05 | 1 | -0.99998 | -0.99991 | 0.99999 |
| 1.926383e-05 | 1 | -0.99993 | -1.0001 | 1 |
| 2.031645e-05 | 1 | -1.0003 | -1.0002 | 1 |
| 2.097353e-05 | 1 | -1 | -0.99986 | 0.99988 |
| 2.233195e-05 | 1 | -0.99994 | -0.9999 | 1.0001 |
| 2.493212e-05 | 1 | -1 | -0.99991 | 1 |
| 2.511214e-05 | 1 | -0.99994 | -0.99988 | 0.99988 |
| 2.708247e-05 | 1 | -0.99991 | -0.99988 | 0.99987 |
| 2.737077e-05 | 1 | -1 | -1 | 1.0001 |
| 3.145850e-05 | 1 | -0.99993 | -0.99995 | 0.99994 |
| 3.335238e-05 | 1 | -1.0001 | -0.99999 | 0.99993 |
| 3.367505e-05 | 1 | -1 | -1.0001 | 1.0002 |
| 3.443855e-05 | 1 | -0.99996 | -1 | 1 |
| 3.625296e-05 | 1 | -1.0001 | -1.0001 | 1.0002 |
| 3.628892e-05 | 1 | -1.0001 | -1.0001 | 1 |
| 3.784736e-05 | 1 | -1 | -0.99995 | 0.99999 |
| 4.201730e-05 | 1 | -1.0002 | -1.0001 | 1 |
| 4.874307e-05 | 1 | -1 | -0.99993 | 1 |
| 4.996208e-05 | 1 | -0.99983 | -0.99999 | 1 |
| 5.204527e-05 | 1 | -0.99994 | -1.0001 | 1.0001 |